# High Voltage (~2 kV) field-plated Al$_{0.64}$Ga$_{0.36}$N-channel HEMTs

Md Tahmidul Alam[1], Jiahao Chen[1], Kenneth Stephenson[2], Md Abdullah-Al Mamun[2], Abdullah Al Mamun Mazumder[2], Shubhra S. Pasayat[1], Asif Khan[2], Chirag Gupta[1]

[1]*Department of Electrical and Computer Engineering, University of Wisconsin-Madison, WI 53706, USA*
[2]*Department of Electrical Engineering, University of South Carolina, Columbia, SC 29208, USA*

*Abstract*—High voltage (~2 kV) AlGaN-channel HEMTs were fabricated with 64% Aluminum composition in the channel. The average on-resistance was ~75 Ω. mm (~21 mΩ. cm$^2$) for L$_{GD}$ = 20µm. Breakdown voltage reached >3 kV (tool limit) before passivation however it reduced to ~2 kV after SiN surface passivation and field plates. The apparent high breakdown voltage prior to passivation can possibly be attributed to the field plate effect of the charged trap states of the surface. The breakdown voltage and R$_{ON}$ demonstrated a strong linear correlation in a scattered plot with ~50 measured transistors. In pulsed IV measurements with 100µs pulse width and 40 V of off-state bias (tool limit), the dynamic R$_{ON}$ increased by ~5% compared to DC R$_{ON}$ and current collapse was <10%.

## I. INTRODUCTION

Wide bandgap results in high critical field in a material therefore wide bandgap semiconductors such as GaN (3.4eV) and SiC are popular as power electronic materials [1]-[3]. The bandgap and critical field of GaN can be further increased by alloying it with AlN (6eV) thus making ternary semiconductors as Al$_x$G$_{1-x}$N [4]-[6]. Higher Al-composition AlGaN is expected to have larger critical field and blocking voltage in power HEMTs (High Electron Mobility Transistors) due the larger bandgap [7], [8]. However, fabrication of high-Al composition AlGaN-channel HEMT is challenging due to difficulty in making good ohmic contacts and low electron mobility leading to high on-resistance (R$_{ON}$) [9], [10]. Grading the AlGaN-channel (from-high-to-low Al-composition) can facilitate the implementation of reasonable ohmic contact resistance [11], [12]. The on-resistance can be improved by doping the barrier to some extent [13].

Previous attempts of making high-voltage (>1kV) high-Al composition (>40%)) AlGaN-channel HEMT with annealed ohmic contact include- demonstration of 1500V Al$_{0.60}$Ga$_{0.4}$N-channel HEMT (peak current density of ~10 mA/mm) by D. Khachariya et. al. and 1800V Al$_{0.51}$Ga$_{0.49}$N-channel HEMT (peak current density ~28 mA/mm) by H. Tokuda et. al. with [14],[15]. R. Maeda et. al. fabricated 1635V Al$_{0.50}$Ga$_{0.50}$N-channel HEMTs with a maximum current density of 250 mA/mm using a more complicated regrown ohmic contact process [16]. However, these works do not include the passivation of the barrier surface and dynamic response of the transistor. Unpassivated barrier surface can potentially have numerous trap states that can impact the dynamic behavior of the transistor. Moreover, the charged trapped states can behave like field plates that can potentially overstate the breakdown voltage. In this work, we have fabricated 64% AlGaN-channel HEMTs (~165 mA/mm peak current density) on sapphire substrate with annealed ohmic contacts that include device passivation and field plates. The breakdown voltage of the fabricated transistors was >3kV before passivation, however it dropped to ~2kV after passivation and field plate inclusion. We achieved decent dynamic properties after surface passivation- the dynamic R$_{ON}$ was <5% higher than DC R$_{ON}$ with 100µs pulse width and 40V switching.

## II. TRANSISTOR FABRICATION AND MEASUREMENT

The structure of the fabricated device is shown in Fig. 1. The epitaxial structure consists of graded n-Al$_x$Ga$_{1-x}$N (30nm, x= 0.30-0.87)/n-Al$_{0.87}$Ga$_{0.13}$N barrier (~20nm, N$_D$ = ~10$^{19}$ cm$^{-3}$)/ Al$_{0.64}$Ga$_{0.36}$N channel (100nm)/ Al$_{0.87}$Ga$_{0.13}$N (140nm) / AlN (260nm) grown on sapphire substrate by metal organic chemical vapor deposition (MOCVD). Device fabrication started with solvent cleaning (3 minutes Acetone, 3 minutes IPA and 1 minute water with ultrasonic power) and subsequent ohmic lithography. Zr/Al/Mo/Au (15/100/40/30) nm was deposited as ohmic metal stacks then was annealed at 950°C for 30 seconds in N$_2$ environment. After that, the graded AlGaN layer was etched up to the ~20nm barrier in the source/drain access region in ICP metal plasma etcher using Cl$_2$. Then, the devices were isolated by a Mesa etch. Following that Ni/Au (100/100) nm gates were deposited in two phases, in each phase the plane of the transistor was inclined by ~30° with respect to the horizontal plane to ensure side-wall coverage of the gate metal. Thereafter, 320nm PECVD Si$_3$N$_4$ was deposited for surface passivation. After that, two field plate trenches were etched in the Si$_3$N$_4$ layer with variable field plate lengths (between 1µm to 3.5µm). Then, bond pad was etched at the metal contacts. Thereafter, field plate metal Ni/Au (200/200) nm was deposited by e-beam

The authors gratefully acknowledge partial support of this research from NSF ASCENT (award number ECCS 2328137), AFOSR under award no. FA9550-23-1-0501 (Program Manager: Dr. Kenneth Goretta) and ARPA-E Award DE-AR0001825.

M.T. Alam, J. Chen, S.S. Pasayat, and C. Gupta are with the Department of Electrical and Computer Engineering, University Wisconsin-Madison at Madison, WI 53706, USA (e-mail: malam9@wisc.edu).

K. Stephenson, M. A. Mamun, A. A. M. Mazumder, A. Khan are with the Department of Electrical Engineering, University of South Carolina, Columbia, SC 29208, USA.

evaporation to fill up the filed plate trenches. The field plates were source-connected. Gate-to-source length ($L_{GS}$) was 2μm, gate length ($L_G$) was 2μm with three variations of gate-drain distance ($L_{GD}$) = 8μm, 15μm and 20μm. The dielectric thickness beneath the field plates were constant- 120nm and 270nm of $Si_3N_4$ under the first and second field plate respectively. More detailed dimensions of the transistor are listed in Table-I.

DC and pulsed measurements were done in Keysight's B1505A SMU (source-measurement unit). During breakdown voltage the transistors were covered with flourinert FC-40 to prevent air breakdown.

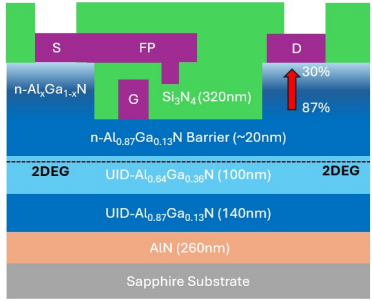

**Fig.1.** Cross-section of the 64% AlGaN-channel HEMT.

TABLE I

| Parameter | Description | Value |
|---|---|---|
| $L_G$ | Gate length | 2μm |
| $L_{GS}$ | Gate to source distance | 2μm |
| $L_{GD}$ | Gate to drain distance | 8,15 and 20μm |
| $L_{FP1}$ | First field plate length | Variable |
| $L_{FP2}$ | Second field plate length | Variable |
| $L_{GF}$ | Gate to field plate distance | 1μm |
| W | Width | 100μm |
| $T_{FP1}$ | Dielectric under first field plate | 120 nm |
| $T_{FP2}$ | Dielectric under second field plate | 270nm |

### III. RESULTS AND DISCUSSION

#### A. IV Characteristics

The average $R_{ON}$ of transistors with $L_{GD}$ = 8μm, 15μm and 20μm were 60Ω.mm (~9.5 mΩ $cm^2$), ~68Ω.mm (~15.5 mΩ $cm^2$) and ~75Ω.mm (~21 mΩ $cm^2$) respectively. The specific on-resistance in mΩ.$cm^2$ was calculated by multiplying the $R_{ON}$ by total channel length ($L_{SD}$ + 2$L_T$). The contact resistance ($R_c$), sheet resistance ($R_{SH}$) and transfer length (2$L_T$) were 4.7 Ω.mm, 2282 Ω/□ and 4μm respectively from TLM measurements [17]. Fig. 2(a) shows the IV curves of a transistor with $L_{GD}$ =20μm.

The transfer curves of some "identical" transistors with $L_{GD}$ = 20μm is shown in Fig. 2(b). The figure shows that the threshold voltage ($V_{TH}$) varies between -8V to -6.5V (assuming 1mA/mm to be the threshold current) with a variation of ~1.5V. This difference in $V_{TH}$ might be due to the non-uniform distribution of Al composition or doping of the graded barrier layer in the source/drain access region, or, due to the non-uniform etch rate of the same layer in $Cl_2$. The average on/off ratio was >$10^4$ (limited by the tool noise).

#### B. Breakdown Voltage

The breakdown voltages of the HEMTs were measured before and after surface passivation. Fig. 3 shows the breakdown measurements before passivation, breakdown voltages were 1380V, 1950V and >3kV (tool limit = 3 kV) for $L_{GD}$ = 8μm, 15μm and 20μm respectively. However, it reduced to <2kV after surface passivation and field plate deposition (Fig. 4). High breakdown voltage before surface passivation can be attributed to the charged surface states acting as field plates. These charged surface states in unpassivated HEMTs can potentially impact the dynamic or switching behavior of the transistor [18]-[20]. Therefore, the breakdown voltage after passivation were taken as the true breakdown. In the passivated HEMTs decent dynamic properties were achieved, it will be discussed in the next section. The breakdown field was ~100V/μm (>150V/μm without passivation) for $L_{GD}$ = 20μm HEMT. For comparison breakdown field (without passivation) was 120V/μm with $L_{GD}$ = 15μm [15], 167V/μm with $L_{GD}$ = 9μm [14] and 300V/μm with $L_{GD}$ = 5.4μm [16]. These results indicate that breakdown field enhancement becomes potentially more challenging with increasing channel length or gate-drain distance. Therefore, the design of field plates number and geometry might play a crucial role in enhancing the breakdown voltage or field in devices in longer $L_{GD}$ transistors. Additionally, surface passivation with high-k dielectrics might help accomplish high breakdown voltage transistors as well [21], [22]. Our future studies will focus on this optimization.

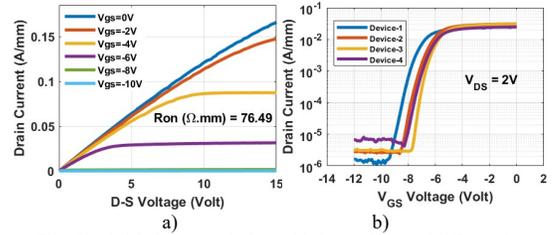

**Fig. 2.** a) IV characteristics with $L_{GD}$= 20μm. b) Transfer curves of some "identical" transistors with $L_{GD}$ =20μm.

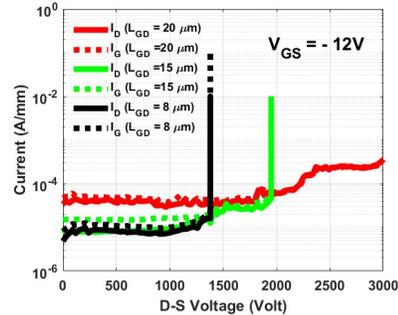

**Fig.3**. Breakdown measurements of transistors with $L_{GD}$ = 8μm, 15μm and 20μm before passivation.

Breakdown voltage and on-resistance were measured in ~50 transistors with different field plate lengths as shown in Fig. 5. The scatter plot indicates a strong linear correlation between the breakdown voltage and on-resistance and also suggests that field plate design can play a vital role in optimizing the performance of the HEMTs in BV-$R_{ON}$ benchmark.

#### C. Preliminary Dynamic Response

The dynamic properties of the field plated HEMTs were characterized by pulsed IV measurements with 100μs pulse width as shown in Fig. 6. In these measurements $V_{GS,Q}$ was set as -12V and $V_{DS,Q}$ was 40V (maximum limit of the tool). For $V_{GS} = 0V$, the dynamic $R_{ON}$ was only ~5% higher than the DC $R_{ON}$ (at $V_{DS} = 1V$) and the amount of current collapse was <10%, indicating decent switching performance of the transistors. We believe the appropriate cleaning of the surface and $Si_3N_4$ passivation neutralized the trap states therefore reasonable dynamic properties were achieved.

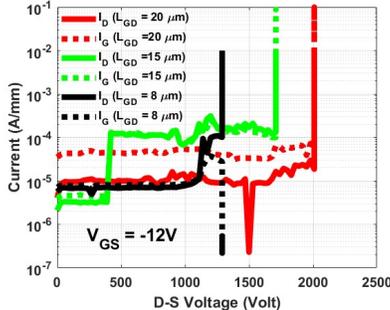

**Fig. 4.** Breakdown measurements with $L_{GD}$ = 8µm, 15µm and 20µm after passivation and field plate deposition. Breakdown voltage reduced in all cases.

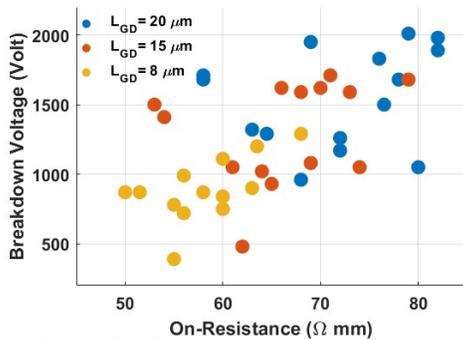

**Fig. 5.** Scatter plot of breakdown voltage vs. $R_{ON}$ indicating a strong linear correlation.

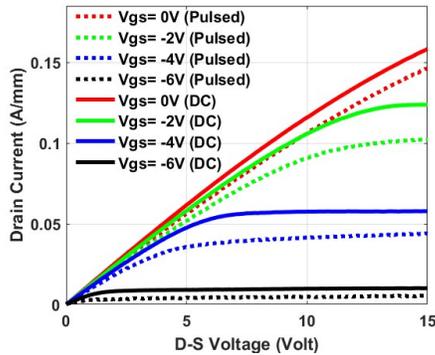

**Fig. 6.** Pulsed and DC IV curves of a HEMT with $L_{GD} = 20$µm ($V_{GSQ} = -12V$, $V_{DSQ} = 40V$, PW=100µs).

## IV. CONCLUSION

High Aluminum composition (64%) AlGaN-channel HEMTs were grown and fabricated for potential high-voltage applications. The fabricated HEMTs reached 3kV (tool limit) before surface passivation, however the breakdown voltage decreased to ~2kV after passivation and field plates deposition ($L_{GD} = 20$µm and $R_{ON} = 75$ Ω.mm or 21 mΩ.$cm^2$). The breakdown voltage after passivation was taken as the true breakdown voltage since unpassivated HEMTs can potentially have charged surface states leading to overstatement of breakdown voltage and poor dynamic performance. The breakdown voltage and $R_{ON}$ showed a strong linear correlation with a little variation due to variation in field plate lengths. The dynamic properties of the HEMTs in pulsed IV measurements with 100µs pulse width were promising, showing only 5% increase in $R_{ON}$ and <10% current collapse with 40V switching (maximum limit of the tool).

REFERENCES

[1] M. Shur, "Wide band gap semiconductor technology: State-of-the-art," *Solid-State Electronics*, vol. 155, pp. 65–75, May 2019, doi: 10.1016/j.sse.2019.03.020.
[2] F. Roccaforte *et al.*, "Emerging trends in wide band gap semiconductors (SiC and GaN) technology for power devices," *Microelectronic Engineering*, vol. 187–188, pp. 66–77, Feb. 2018, doi: 10.1016/j.mee.2017.11.021.
[3] T. J. Flack, B. N. Pushpakaran, and S. B. Bayne, "GaN Technology for Power Electronic Applications: A Review," *J. Electron. Mater.*, vol. 45, no. 6, pp. 2673–2682, Jun. 2016, doi: 10.1007/s11664-016-4435-3.
[4] R. J. Kaplar *et al.*, "Review—Ultra-Wide-Bandgap AlGaN Power Electronic Devices," *ECS J. Solid State Sci. Technol.*, vol. 6, no. 2, p. Q3061, Dec. 2016, doi: 10.1149/2.0111702jss.
[5] A. A. Shuvo, Md. R. Islam, and Md. T. Hasan, "Ultrawide-bandgap AlGaN-based HEMTs for high-power switching," *J Comput Electron*, vol. 19, no. 3, pp. 1100–1106, Sep. 2020, doi: 10.1007/s10825-020-01532-3.
[6] J. Y. Tsao *et al.*, "Ultrawide-Bandgap Semiconductors: Research Opportunities and Challenges," *Advanced Electronic Materials*, vol. 4, no. 1, p. 1600501, 2018, doi: 10.1002/aelm.201600501.
[7] T. Nanjo *et al.*, "AlGaN Channel HEMT With Extremely High Breakdown Voltage," *IEEE Transactions on Electron Devices*, vol. 60, no. 3, pp. 1046–1053, Mar. 2013, doi: 10.1109/TED.2012.2233742.
[8] H. Tokuda *et al.*, "High Al Composition AlGaN-Channel High-Electron-Mobility Transistor on AlN Substrate," *Appl. Phys. Express*, vol. 3, no. 12, p. 121003, Dec. 2010, doi: 10.1143/APEX.3.121003.
[9] H. Xue *et al.*, "All MOCVD grown Al0.7Ga0.3N/Al0.5Ga0.5N HFET: An approach to make ohmic contacts to Al-rich AlGaN channel transistors," *Solid-State Electronics*, vol. 164, p. 107696, Feb. 2020, doi: 10.1016/j.sse.2019.107696.
[10] K. Hussain *et al.*, "High figure of merit extreme bandgap Al $_{0.87}$ Ga $_{0.13}$ N-Al $_{0.64}$ Ga $_{0.36}$ N heterostructures over bulk AlN substrates," *Appl. Phys. Express*, vol. 16, no. 1, p. 014005, Jan. 2023, doi: 10.35848/1882-0786/acb487.
[11] H. Ye, M. Gaevski, G. Simin, A. Khan, and P. Fay, "Electron mobility and velocity in Al0.45Ga0.55N-channel ultra-wide bandgap HEMTs at high temperatures for RF power applications," *Applied Physics Letters*, vol. 120, no. 10, p. 103505, Mar. 2022, doi: 10.1063/5.0084022.
[12] M. Gaevski *et al.*, "Ultrawide bandgap Al $_x$ Ga $_{1-x}$ N channel heterostructure field transistors with drain currents exceeding 1.3 A mm $^{-1}$," *Appl. Phys. Express*, vol. 13, no. 9, p. 094002, Sep. 2020, doi: 10.35848/1882-0786/abb1c8.
[13] H. Xue *et al.*, "High-Current-Density Enhancement-Mode Ultrawide-Bandgap AlGaN Channel Metal–Insulator–Semiconductor Heterojunction Field-Effect Transistors with a Threshold Voltage of 5 V," *physica status solidi (RRL) – Rapid Research Letters*, vol. 15, no. 6, p. 2000576, 2021, doi: 10.1002/pssr.202000576.
[14] D. Khachariya *et al.*, "Record >10 MV/cm mesa breakdown fields in Al0.85Ga0.15N/Al0.6Ga0.4N high electron mobility transistors on native AlN substrates," *Applied Physics Letters*, vol. 120, no. 17, p. 172106, Apr. 2022, doi: 10.1063/5.0083966.
[15] H. Tokuda *et al.*, "High Al Composition AlGaN-Channel High-Electron-Mobility Transistor on AlN Substrate," *Appl. Phys. Express*, vol. 3, no. 12, p. 121003, Dec. 2010, doi: 10.1143/APEX.3.121003.
[16] R. Maeda, K. Ueno, A. Kobayashi, and H. Fujioka, "AlN/Al $_{0.5}$ Ga $_{0.5}$ N HEMTs with heavily Si-doped degenerate GaN contacts prepared via pulsed sputtering," *Appl. Phys. Express*, vol. 15, no. 3, p. 031002, Mar. 2022, doi: 10.35848/1882-0786/ac4fcf.
[17] A. Mamun *et al.*, "Al $_{0.64}$ Ga $_{0.36}$ N channel MOSHFET on single crystal bulk AlN substrate," *Appl. Phys. Express*, vol. 16, no. 6, p. 061001, Jun. 2023, doi: 10.35848/1882-0786/acd5a4.
[18] Y. Dora, A. Chakraborty, L. Mccarthy, S. Keller, S. P. Denbaars, and U. K. Mishra, "High Breakdown Voltage Achieved on AlGaN/GaN

HEMTs With Integrated Slant Field Plates," *IEEE Electron Device Letters*, vol. 27, no. 9, pp. 713–715, Sep. 2006, doi: 10.1109/LED.2006.881020.
[19] S. Arulkumaran, T. Egawa, H. Ishikawa, T. Jimbo, and Y. Sano, "Surface passivation effects on AlGaN/GaN high-electron-mobility transistors with SiO2, Si3N4, and silicon oxynitride," *Applied Physics Letters*, vol. 84, no. 4, pp. 613–615, Jan. 2004, doi: 10.1063/1.1642276.
[20] S. Arulkumaran, "Surface passivation effects in AlGaN/GaN HEMTs on high-resistivity Si substrate," in *2007 International Workshop on Physics of Semiconductor Devices*, Dec. 2007, pp. 317–322. doi: 10.1109/IWPSD.2007.4472507.
[21] M. W. Rahman, N. K. Kalarickal, H. Lee, T. Razzak, and S. Rajan, "Integration of high permittivity BaTiO3 with AlGaN/GaN for near-theoretical breakdown field kV-class transistors," *Applied Physics Letters*, vol. 119, no. 19, p. 193501, Nov. 2021, doi: 10.1063/5.0070665.
[22] T. Razzak *et al.*, "BaTiO3/Al0.58Ga0.42N lateral heterojunction diodes with breakdown field exceeding 8 MV/cm," *Applied Physics Letters*, vol. 116, no. 2, p. 023507, Jan. 2020, doi: 10.1063/1.5130590.